\documentclass[pdflatex]{sn-jnl}
\usepackage{graphicx}%
\usepackage{multirow}%
\usepackage{amsmath,amssymb,amsfonts}%
\usepackage{amsthm}%
\usepackage{mathrsfs}%
\usepackage[title]{appendix}%
\usepackage{xcolor}%
\usepackage{textcomp}%
\usepackage{manyfoot}%
\usepackage{booktabs}%
\usepackage[authoryear,round]{natbib}
\usepackage{algorithm}%
\usepackage{algorithmicx}%
\usepackage{algpseudocode}%
\usepackage{listings}%

\theoremstyle{thmstyleone}%

\theoremstyle{thmstyletwo}%

\theoremstyle{thmstylethree}%

\begin{document}

\title[Habitability Study of Terrestrial Planets: Application to Venus-like Worlds]{Habitability Study of Terrestrial Planets: Application to Venus-like Worlds}

\author[1]{\fnm{Swathi} \sur{Raviprakash}}

\author*[1]{\fnm{Madhu Kashyap} \sur{Jagadeesh}}\email{madhu.kashyap@sju.edu.in}

\author[2,3]{\fnm{Margarita} \sur{Safonova}}

\author[3]{\fnm{Oleg} \sur{Kotsyurbenko}}

\affil*[1]{\orgdiv{Department of Physics}, 
\orgname{St Joseph's University}, 
\orgaddress{\city{Bengaluru}, \state{Karnataka}, \country{India}}}

\affil[2]{\orgname{M. P. Birla Institute of Fundamental Research}, 
\orgaddress{\city{Bengaluru}, \state{Karnataka}, \country{India}}}

\affil[3]{\orgname{NoRCEL Institute}, 
\orgaddress{\city{Leeds}, \country{United Kingdom}}}

\abstract{The study of planetary habitability beyond Earth remains central and a challenging project in planetary science. Analysis of large volumes of planetary data from space missions such as CoRoT, Kepler, and JWST is directed ultimately at finding a planet similar to Earth—the Earth’s twin, and answering the question of potential exo-habitability. The Earth Similarity Index (ESI) is a first step in this quest, ranging from 1 (Earth) to 0 (totally dissimilar to Earth). To identify planets that may be habitable to the extreme forms of life, we introduced the Mars Similarity Index (MSI). However, extreme forms of life have also been hypothesized under specific conditions in the upper atmosphere of Venus, motivating comparative habitability studies beyond Earth and Mars.
The Venus Similarity Index (VSI) introduced here is defined as the geometric mean of radius, density, escape velocity, and surface temperature, normalized in Venus Units (VU). VSI values range from 0 (complete dissimilarity) to 1 (maximum similarity). The VSI provides a comparative framework for identifying Venus-like planetary environments within exoplanet populations. To explore habitability evolution, we further introduce the Ancient Venus Similarity Index (AVSI) and the Future Earth Similarity Index (FESI) to examine early Venusian conditions relative to ancient Earth and to assess potential future evolutionary pathways for Earth-like planets.}

\keywords{Venus, Habitability and Exoplanets}

\maketitle
\vspace{-1em}

\section{Introduction}
\label{sec1}

Answering questions about the origin and evolution of life and search for it in the Universe has been the primary motivation of astrobiologists and planetary scientists. The discovery of 51~Pegasi~b by \citet{Mayor1995AStar} marked the first detection of an exoplanet orbiting a Sun-like star. Currently, 6500+ exoplanets have been detected using various space and ground-based observatories (such as Kepler, JWST, TESS, HARPS, etc.)\footnote{NASA Exoplanet Archive, \url{https://exoplanetarchive.ipac.caltech.edu/}}. One of the central aims of contemporary exoplanet research is the identification of Earth‑like worlds—often termed “Earth 2.0”—and the assessment of their potential exo‑habitability. However, a comprehensive assessment of exoplanet's habitability requires highly detailed information, which is currently beyond our reach. For now, our best approach is to compare the measurable (or inferable) properties of planets with Earth to determine their potential habitability. In this context, we define habitability in terms of Earth-like conditions, or some modified but still recognizable form. 

While Earth serves as the standard for habitability (being the only known inhabited planet), planets such as Mars and Venus have long intrigued scientists due to their similarities with Earth. However, despite their proximity to Earth, current Mars and Venus exhibit different environmental conditions \citep{Jakosky2001MarsHistory, Taylor2018Venus:Planet}.  

Venus, Earth, and Mars represent divergent evolutionary pathways of terrestrial planets, shaped by complex interactions between their interiors, atmospheres, and stellar environments, providing key insights into planetary habitability and climate evolution \citep{Jakosky2025UsingEvolution}. Here, the basic criteria and data are already available, using which we can start gauging the potential habitability of exoplanets. The challenge here is to determine which exoplanet parameter(s) are important in finding this similarity. Several assessment scales were introduced, beginning with the Earth Similarity Index (ESI) \citep{Schulze-Makuch2011AExoplanets}, to indicate how similar to Earth is an exoplanet to judge its habitability potential. 

Several studies indicate that ancient Mars may have supported environments conducive to life \citep{Abramov2016ThermalMars, Valdivia-Silva2013NoachianLife}, thereby motivating continued investigations into the planet's biological potential. We have extended the ESI to the Mars Similarity Index (MSI) to search for 
Mars-like exoplanets as potential hosts for extremophile life forms \citep{KashyapJagadeesh2017IndexingWorlds}.
In addition, studies that assess the evolution of temperature and surface conditions on Venus also suggest that young Venus-like exoplanets could be potential targets in the search of extra-terrestrial life \citep{Way2016WasSystem}. 
Current Venus has a surface pressure of about 90 bars and a surface temperature of about $462^{\circ}$C, while the critical temperatures are $374^{\circ}$C for pure water, and about $410^{\circ}$C for salty seawater. Exoplanets with surface temperatures below critical might host stable liquid water conditions under suitable surface pressure conditions. Therefore, Venus-like exoplanets with surface temperatures below, say $150^{\circ}$C, are of biological interest. 
Therefore, in this paper, we introduce a new metric tool, the Venus Similarity Index (VSI) to search for Venus-like exoplanets as well as the Ancient Venus Similarity Index (AVSI), to find ancient Venus-like exoplanets in the data, which can have conditions similar to those on Earth and favourable for life.  

\begin{figure}
\hspace{-1.2in}
\includegraphics[width=0.8\linewidth]{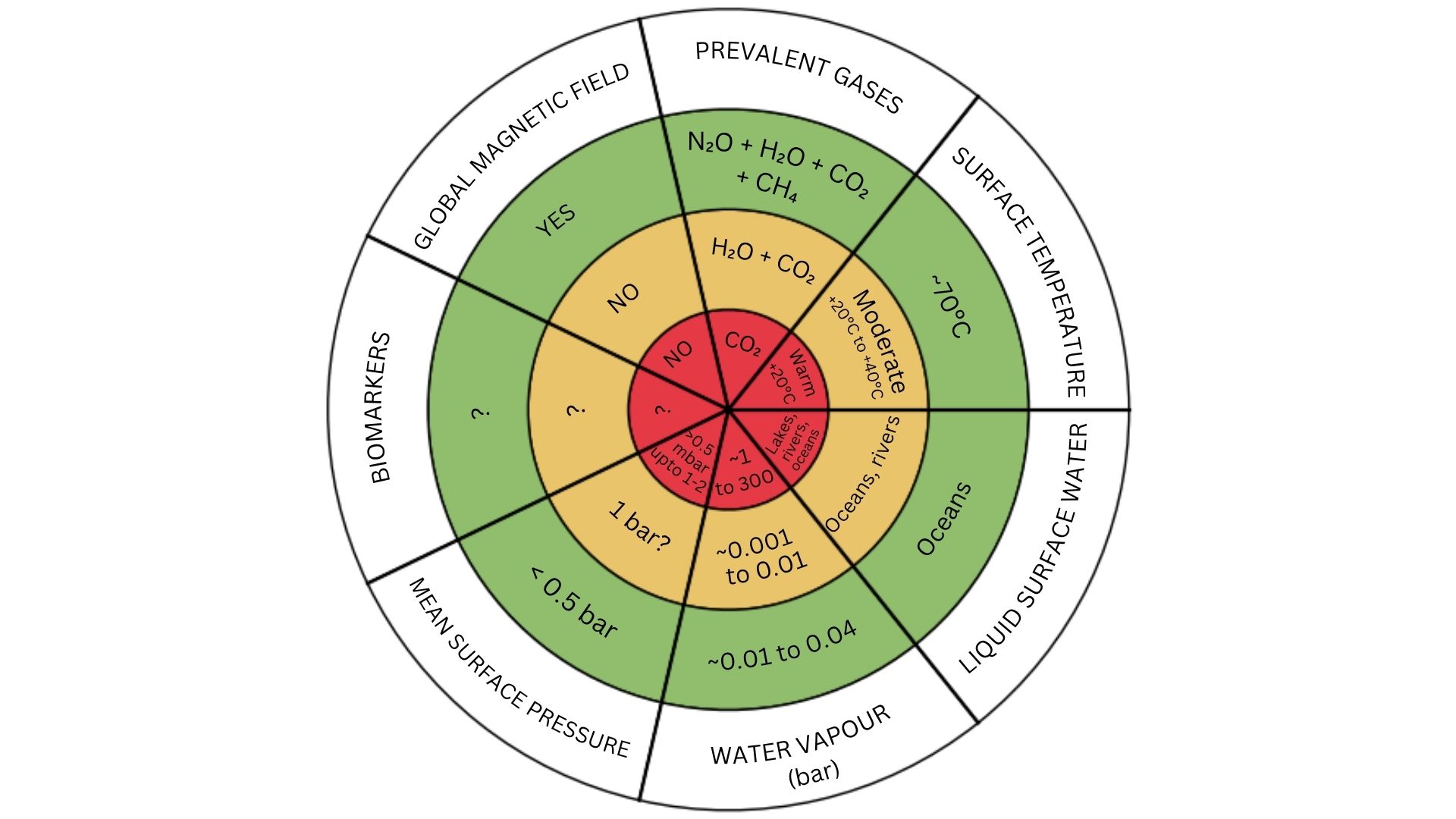}
\hspace{-0.9in}
\includegraphics[width=0.8\linewidth]{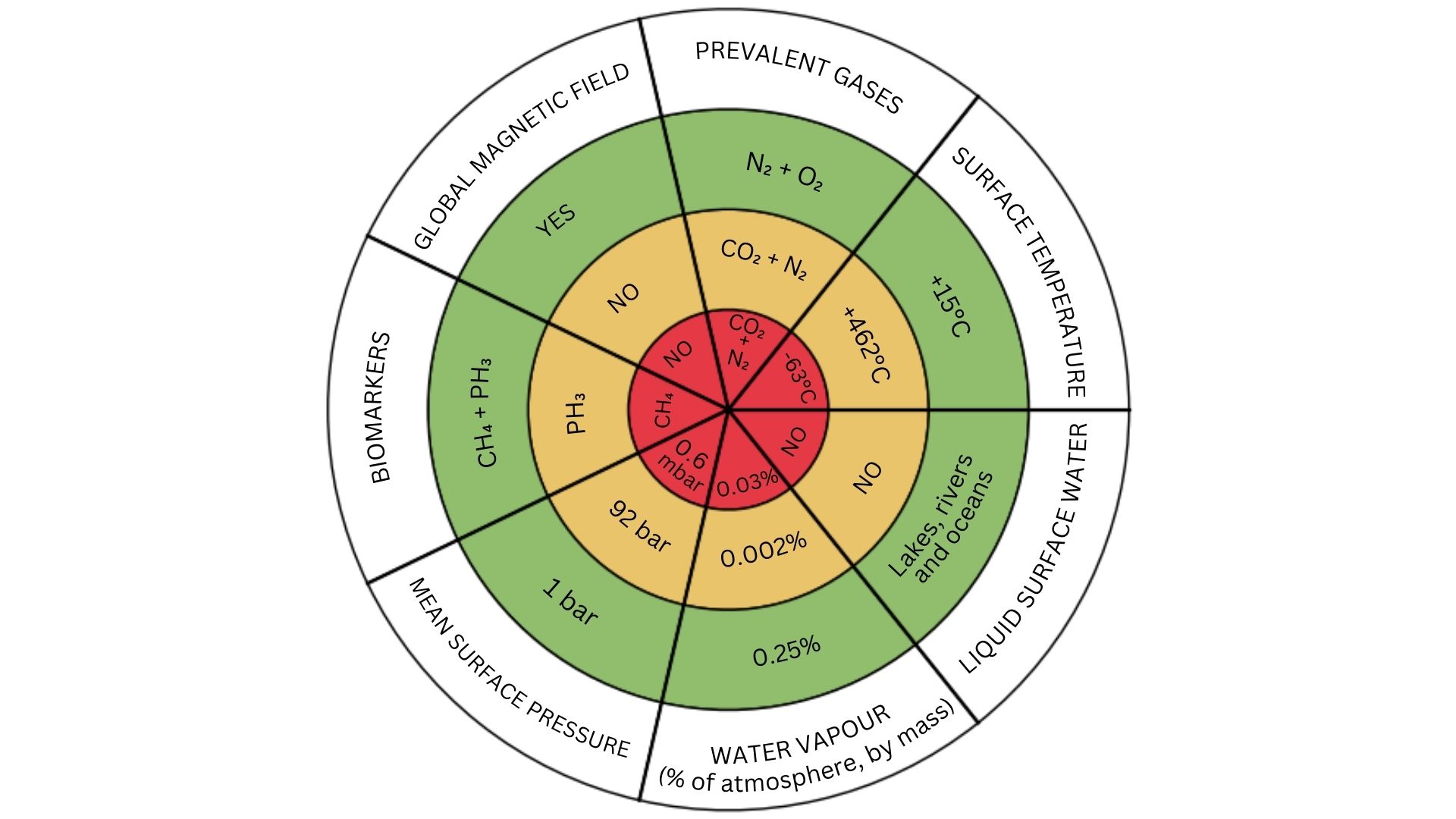}
\caption{Habitability factors at time period from $\sim$4.1 Ga till end of LHB (left) and current (right) Earth (green), Venus (yellow), and Mars (red). } 
\label{current pie plot}
\end{figure}

Ancient Venus could have supported habitable conditions (see Fig.~\ref{current pie plot}) for nearly 3 billion years \citep{Way2020VenusianExoplanets}. Ancient Venus may have had Earth-like conditions, including liquid water oceans, land–ocean interfaces, and favourable chemical and energy environments that could have supported the origin of life.
Venus may even have been more habitable than Earth in the beginning with surface water including oceans, land–ocean interfaces, and favorable chemical and energy environments that could have supported the origin of life, like moderate temperatures and even a global magnetic field \citep{Safonova2021PlanetaryLate}.
Anyway, even after surface conditions became hostile, life might have persisted within the planet’s cloud layers where temperature and pressure conditions are similar to those on the earth’s surface at some altitudes, so the possibility of extant life remains a serious hypothesis \citep{Izenberg2021TheEquation}. Hence, we introduce the Ancient Venus Similarity Index (AVSI) metric scale to find ancient Venus-like exoplanets in the data. To place habitability within an evolutionary context, we extend our framework by introducing the Future Earth Similarity Index (FESI), which represents the declining phase of Earth-like planets as their host stars evolve toward the red giant stage. In contrast to AVSI, which captures early Venus-like potentially habitable conditions, FESI enables us to explore the terminal stages of habitability. Together, these indices define a continuous trajectory of planetary habitability spanning its initial stage, peak, and eventual loss.

The structure of this paper is as follows. In Section~\ref{sec2}, we discuss the Earth and Mars similarity indices (ESI and MSI). In Section \ref{sec3}, we present the VSI formulation, and in Section~\ref{sec4.1}, we discuss the AVSI and FESI.
Finally, discussion and conclusions are in Section~\ref{sec7}. 

\section{Earth Similarity Index and Mars Similarity Index}\label{sec2}

Distance or similarity measures are commonly used in tasks such as object classification, clustering, information retrieval, and evaluating the degree of overlap between quantitative datasets. Similarity indices are effective tools for organizing and identifying patterns within large and complex datasets. These are straightforward to compute and offer a clear quantitative measure of how much a system deviates from a chosen reference state, typically expressed on a scale from 0 to 1. Such indices are widely applied across disciplines, including mathematics, e.g. set theory and fuzzy logic \citep{zadeh1965fuzzy}; ecology, e.g. Sørensen similarity index \citep{sorensen1948method}; computer imaging, e.g. structural similarity index \citep{wang2004ssim}, and chemistry, e.g. Jaccard-Tanimoto similarity index \citep{jaccard1901etude}, among others. The Earth Similarity Index (ESI) was introduced by \citet{Schulze-Makuch2011AExoplanets} and it paved the way for the development of the Mars Similarity Index (MSI) by \citet{KashyapJagadeesh2017IndexingWorlds}, both of which are used to scale the similarity of exoplanets. 
The ESI is an initial metric used to assess how similar a planet is to Earth, with values ranging from 1 (identical to Earth) to 0 (completely dissimilar). It is calculated using four key planetary parameters: radius, density, escape velocity, and surface temperature, where the input parameters are expressed in Earth Units (EU) (Appendix~A Table), except for surface temperature, which is in Kelvin.

The ESI for each physical property is given by
\begin{equation} \label{esi}
ESI_x = \left( 1 - \left|\frac{x - x_0}{x + x_0}\right| \right)^{w_x},
\end{equation}

where \(x\) denotes the planetary property, \(x_0\) is the corresponding reference value for Earth, and \(w_x\) is the weight exponent associated with that parameter. The weight exponents \(w_x\) are determined by adopting a similarity threshold \(V = 0.8\), corresponding to a region of very high similarity, and by defining lower and upper bounds \(x_a\) and \(x_b\) for each parameter such that \(x_a < x_0 < x_b\). The lower and upper weight exponents are given by
\begin{equation}
w_a = \frac{\ln V}{\ln\!\left[1-\left|\frac{x_0 - x_a}{x_0 + x_a}\right|\right]}, \quad
w_b = \frac{\ln V}{\ln\!\left[1-\left|\frac{x_0 - x_b}{x_0 + x_b}\right|\right]}\,.
\label{eq:weights}
\end{equation}\\

The final weight exponent is then obtained as the geometric mean
\begin{equation}
w_x = \sqrt{w_a \times w_b} \,.
\label{eq:weight}
\end{equation}

The global ESI is the geometrical mean of interior ESI (which depends upon radius and density) and the surface ESI (which depends on the escape velocity and surface temperature),

\begin{equation}
ESI = \left( ESI_{\mathrm{interior}} \times ESI_{\mathrm{surface}} \right)^{1/2}.
\end{equation}\\

\begin{figure}
\centering
\includegraphics[width=0.7\linewidth]{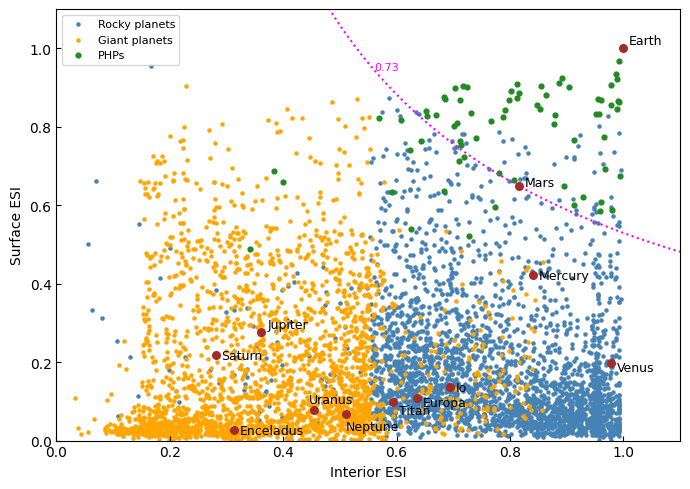}
\caption{Plot of Interior ESI vs Surface ESI. Yellow dots represent \emph{giant planets}, blue dots represent \emph{rocky planets}, brown dots represent \emph{Solar System objects} and green dots represent \emph{Potentially Habitable Planets} (PHPs). The dotted line represents the ESI value of Mars (0.73). We have found 103 planets above this threshold.}
\label{ESIscatter}
\end{figure}

\begin{figure}
\centering
\includegraphics[width=0.7\linewidth]{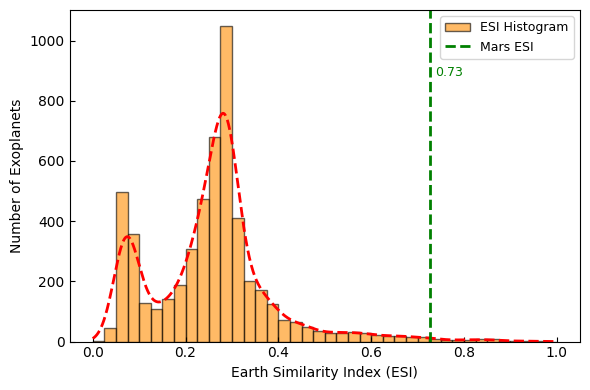}
\caption{Histogram representation of total exoplanets to their corresponding ESI values. The red dotted line represents the best fit to bimodal Gaussian curve to the histogram and the green dotted line represents the ESI of Mars = 0.73.}
\label{ESIhist}
\end{figure}

In our calculations, the exoplanet data is obtained from the online Exoplanets Catalog (PHL-EC) maintained by the Planetary Habitability Laboratory (PHL) at the University of Puerto Rico, Arecibo\footnote{\url{https://phl.upr.edu/hwc/data}}. The catalog contains 5600 exoplanets (as of September 2025). PHL uses the ESI index to estimate the potential habitability of all known exoplanets: planets with an ESI of 0.8 or higher are considered Earth-like, while those with an ESI of 0.73 or higher (such as Mars) are optimistically labeled as potentially habitable planets (PHP). At the time of writing (PHL-EC data update was of January 2024), there are 70 such assumed PHPs in PHL-EC. Fig.~\ref{ESIscatter} shows the scatter plot of interior versus surface ESI. Considering ESI of Mars (0.73) as the threshold, we have found 103 planets above this threshold. Of these 103 planets, 56 are registered in PHL-EC as PHPs, additionally, we have found 47 potentially habitable candidates. PHPs of PHL-EC under threshold are found to have masses between 4 to 36~EU and surface temperatures below $\sim$250~K. The range of these parameters affects the global ESI value and pushes the PHPs below the defined threshold.
The global ESI distribution is presented as a histogram in Fig.~\ref{ESIhist}, which exhibits two prominent peaks and approximates a bimodal Gaussian distribution in the initial range of the plot.

Mars often has captured interests of planetary scientists and astrobiologists due to the evidence of liquid water \citep{Fairen2010AMars, Read2015TheReview, Pavlov2026DoesMudstone}. The detection of episodic methane variations in the Martian atmosphere by the Curiosity rover \citep{Grotzinger2015DepositionMars} represented a significant advancement in studies of Martian habitability, as methane may be linked to processes relevant to potential microbial activity.
Complementary to the ESI, the Mars Similarity Index (MSI) is defined as the geometric mean of radius, density, surface temperature and escape velocity of planets -- all measured in Mars Units (MU) -- and ranges from 0 (totally dissimilar to Mars) to 1 (identical to Mars).
The MSI follows an identical mathematical formulation:
\begin{equation}
MSI_x = \left( 1 - \left|\frac{x - x_0}{x + x_0}\right| \right)^{w_x},
\end{equation}
where all symbols have the same meaning as in Eq.~\ref{esi}. In this case, all input parameters are expressed in Mars Units (MU), except for surface temperature, which remains in Kelvin. The Mars-like parameter ranges used for calibration are tabulated in Appendix~A table, adopted from \citet{KashyapJagadeesh2017IndexingWorlds}.

\begin{figure}
\centering
\includegraphics[width=0.7\linewidth]{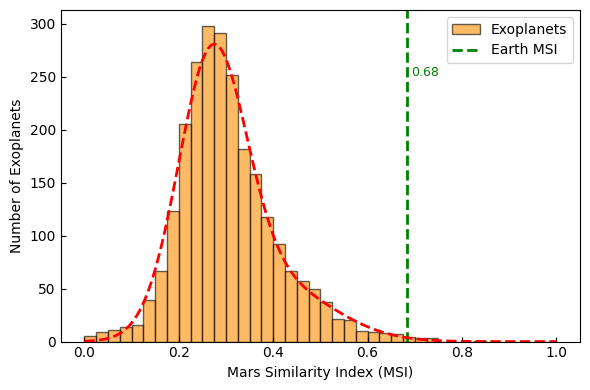}
\caption{Histogram representation of total exoplanets to their corresponding MSI values. The red dotted line represents the best fit of a Gaussian curve to the histogram and the green dotted line represents the MSI of Earth = 0.68 .}
\label{MSIhist}
\end{figure}

\begin{figure}
\centering
\includegraphics[width=0.7\linewidth]{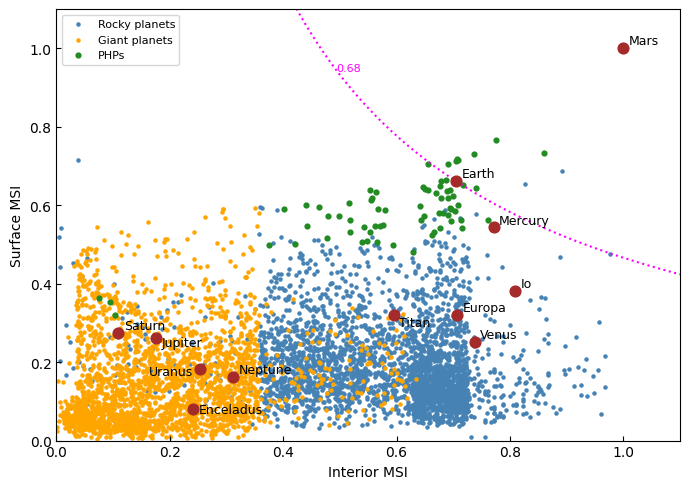}
\caption{Plot of Interior MSI vs Surface MSI. Yellow dots represent \emph{giant planets}, blue dots represent \emph{rocky planets}, brown dots represent \emph{Solar System objects} and green dots represent \emph{Potentially Habitable Planets} (PHPs). The dashed line represents MSI value of Earth (0.68). We have found 11 planets above this threshold.}
\label{MSIscatter}
\end{figure}

The histogram of the global MSI plotted for a sample of 5600 exoplanets is shown in Fig.~\ref{MSIhist}. Fig.~\ref{MSIscatter} presents a scatter plot of the interior versus surface MSI for the same dataset. The dashed curve in the scatter plot represents the adopted threshold, set equal to the MSI value of Earth (0.68), which is used to identify Mars-like exoplanets.

Based on this criterion, we identify 11 Mars-like exoplanets in the current sample, of which 9 are PHPs. Thus, we have found 2 more planets which are Mars-like. These planets may represent environments capable of supporting extremophilic organisms including methane-producing archaea under suitable conditions. The relatively small number of such candidates is likely influenced by observational selection effects, as current telescopes are not designed for the detection of small, rocky planets.


\section{Venus Similarity Index}\label{sec3}

Current Venus, despite being closer to Earth in size and mass, presents extreme surface conditions with temperatures exceeding 730~K and atmospheric pressures 90 times greater than Earth (the current physical properties of Venus are listed in Appendix~B table). 

Venus is the most Earth-like planet in our Solar System, with extremely similar conditions in the past, which later diverged as these planets evolved \citep{Way2016WasSystem, Westall2023TheVenus, Izenberg2021TheEquation}. This implies that Venus may be a natural laboratory for studying the evolution of habitability \citep{Kane2019VenusScience}. 
The recent report of PH$_3$ detection in Venus upper atmosphere have suggested the possibility of life there, as PH$_3$ as an indicator of the microbial life cycle on Earth and thus is a possible biosignature \citep{Greaves2021PhosphineVenus}. 
Re-analysis of data collected by the Pioneer Venus Large Probe from 1978 suggested that clouds of Venus are actually 60\% water \citep{Mogul2025ReAnalysisAerosols}. 

The search for extraterrestrial life has traditionally focused on environments similar to Earth, prioritizing the presence of liquid water on a planetary surface \citep{Kasting1993HabitableStars, Forget2010HabitabilityPlanets}. With its current surface temperatures exceeding 730~K and atmosphere dominated by CO$_2$ resulting from the runaway greenhouse effect \citep{Taylor2018Venus:Planet}, Venus is definitively outside the traditional habitable zone, but it presents a compelling case for exploring habitability in extreme environments.
The unique chemical and thermal profile of the Venusian atmosphere, confirmed by decades of data from missions like Veneras and Magellan \citep{Titov2018CloudsVenus, Way2016WasSystem}, provides a benchmark for environments where life may persist without a terrestrial-like surface. 
The temperate cloud layer of Venus is located between 48 and 60 km in altitude and maintains pressure and temperature regimes ($\sim$290--320~K) within the range where liquid water and known Earth extremophiles could potentially survive \citep{Limaye2018VenusClouds, Kotsyurbenko2021ExobiologyDetection}. The presence of concentrated sulfuric acid aerosols, however, dictates that any potential life must be highly acidophilic \citep{Skladnev2021WatersulfuricVenus} and capable of thriving in high-radiation conditions \citep{Schulze-Makuch2021TheClouds}.

The Venus Similarity Index (VSI) is developed following the same mathematical framework as the ESI and MSI proposed in earlier studies \citep{Schulze-Makuch2011AExoplanets, KashyapJagadeesh2017IndexingWorlds}. The VSI is a metric used to evaluate how closely a planet resembles Venus. Its values range from 1, representing a planet identical to Venus, to 0, indicating a completely dissimilar planet. The VSI is calculated using four fundamental planetary parameters: radius, density, escape velocity, and surface temperature.

\noindent
The global VSI is computed as the geometric mean of the individual similarity components:
\begin{equation}
VSI = \left( VSI_R \times VSI_\rho \times VSI_{V_e} \times VSI_{T_s} \right)^{1/4}.
\end{equation}
By construction, the VSI provides a normalized similarity measure that enables a direct quantitative comparison of planetary parameters relative to Venus within the same formalism as the ESI and MSI.\\
The \emph{interior VSI} is defined as the geometric mean of the similarity components associated with planetary radius and bulk density:
\begin{equation}
VSI_{\mathrm{interior}} = \left( VSI_R \times VSI_\rho \right)^{1/2}.
\end{equation}
The \emph{surface VSI} is defined as the geometric mean of the similarity components associated with escape velocity and surface temperature:
\begin{equation}
VSI_{\mathrm{surface}} = \left( VSI_{V_e} \times VSI_{T_s} \right)^{1/2}.
\end{equation}
The global VSI is then be expressed in terms of these two components as
\begin{equation}
VSI = \left( VSI_{\mathrm{interior}} \times VSI_{\mathrm{surface}} \right)^{1/2}.
\end{equation}

\citet{Kane2026ImagingObservatory} study outlines how the Habitable Worlds Observatory (HWO) will enable the detection and atmospheric characterization of exo-Venuses by combining direct imaging with UV, visible, and spectropolarimetric measurements, revealing key features such as sulfur dioxide absorption and cloud properties that distinguish Venus-like worlds from temperate terrestrial planets.
For surface temperature, the admissible range was not selected directly from Solar System analogues but instead derived from physical and biochemical constraints relevant to planetary habitability and geophysical stability. Specifically, CO$_2$ sublimes at 194.7 K ($-78.5$°C) at Earth atmospheric pressure \citep{Barber1966TheDioxide}. This implies the same effect below 400 K. Condensation of CO$_2$ decreases the lapse rate and hence reduces the magnitude of the greenhouse effect \citep{Kasting1991CO2Mars}. Around 400 K, water can remain liquid under appropriate pressure conditions, enabling potential habitability scenarios \citep{Vladilo2013ThePressure, Kopparapu2014HabitableMass}. Above $\sim$800~K, liquid water is absent, and complex organic molecules are rapidly destroyed \citep{Fang2015ThermalInvestigation, Cockell2016Habitability:Review}. Also, above this range of temperatures, some minerals may undergo thermal decomposition, phase transitions, or dehydration \citep{Hirschmann2000MantleComposition}. Some extremophilic microorganisms on Earth demonstrate survivability up to 395 K, supporting the lower end of threshold in weight exponent calculations \citep{Takai2008CellCultivation}.
A team of scientists finds that while most plants wither or stop growing in extreme heat, \emph{T. oblongifolia} (native to Death Valley) thrives as temperatures climb\footnote{https://www.sciencedaily.com/releases/2025/11/251109032410.htm}. Researchers also found that when exposed to scorched-Earth conditions (over 120°F/49°C), the plant tripled its biomass in just 10 days, while other heat-tolerant species ceased growth entirely. 
\citet{barbier2016thermococcus} highlights the behaviour of the microbe \emph{T. gammatolerans}, which can withstand doses of 30,000 grays -- 6,000 times the lethal limit for humans. It is found near nuclear disaster sites (like Chernobyl), and evolves in an environment with almost no natural ionizing radiation.  
This microbe is found to feed on sulfur compounds and thrives at temperatures around 88°C (190°F). However, radiation resistance didn't seem to be a survival necessity in the microbe's habitat.

Weight exponents were calculated following Eqs.~\ref{eq:weights} and \ref{eq:weight}. The effective temperature range adopted in this study is $300–800$~K. 
This ensures that weight exponent for the temperature reflects both geophysical and biochemical considerations while remaining anchored in Venus’s mean surface temperature of 737 K.

The parameters required for our calculations include the interior parameters, radius \emph{R} and density $\rho$ and the surface parameters, escape velocity $V_e$ and surface temperature $T_S$. All these parameters (except surface temperature) are expressed in units relative to the planet in consideration, i.e. Venus Units (VU) for VSI calculations. For weight exponent calculation of VSI, we adopt Mars as the lower bound and Earth as the upper bound (Eq.~\ref{eq:weights}). The radius limits are adopted to be 0.58 to 1.05 VU, the density range is 0.75 to 1.05 VU and the escape velocity range is 0.48 to 1.08 VU. The normalization values adopted for Venus are provided in Appendix~A table. The exoplanets with the highest VSI values, together with their corresponding surface and interior VSI components, are listed in Table~\ref{tab:top10vsi}.

\begin{table}
\centering
\caption{Exoplanets with highest VSI values}
\label{tab:top10vsi}
\begin{tabular}{lccccccc}
\hline
Names & Radius (VU) & Escape (VU) & Density (VU) & Temp (K) & VSI$_I$ & VSI$_S$ & VSI \\
\hline
Kepler-1650 b     & 1.01 & 1.00 & 0.99 & 758.25 & 0.99 & 0.98 & 0.98 \\
Kepler-399 b      & 1.01 & 1.00 & 0.99 & 704.70 & 0.99 & 0.97 & 0.98 \\
Kepler-1049 b     & 1.00 & 0.99 & 0.99 & 798.94 & 0.99 & 0.95 & 0.97 \\
Kepler-1505 b     & 0.97 & 0.96 & 0.99 & 726.42 & 0.96 & 0.97 & 0.97 \\
Kepler-398 b      & 0.97 & 0.96 & 0.97 & 859.90 & 0.96 & 0.91 & 0.94 \\
Kepler-197 e      & 0.95 & 0.93 & 0.96 & 799.69 & 0.94 & 0.93 & 0.93 \\
Kepler-271 c      & 1.00 & 0.99 & 0.99 & 938.66 & 0.99 & 0.88 & 0.93 \\
EPIC 206032309 b  & 1.06 & 1.07 & 1.03 & 687.43 & 0.93 & 0.93 & 0.93 \\
LHS 475 b         & 1.04 & 1.04 & 1.01 & 625.52 & 0.95 & 0.90 & 0.93 \\
Kepler-1347 b     & 1.08 & 1.10 & 1.03 & 713.52 & 0.91 & 0.94 & 0.92 \\
\hline
\end{tabular}
\end{table}
\begin{figure}
\centering
\includegraphics[width=0.7\linewidth]{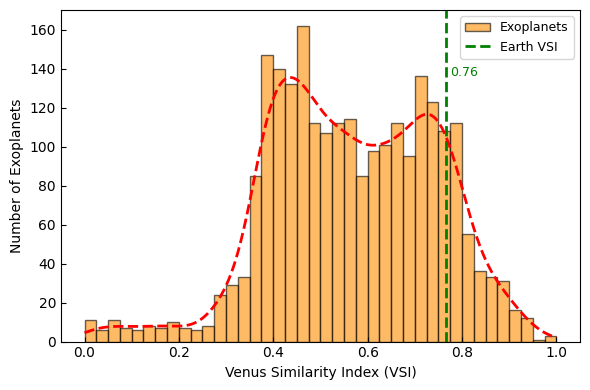}
\caption{Histogram representation of total exoplanets to their corresponding VSI values. The red dotted line represents the fit to bimodal distribution and the green dotted line represents the VSI of Earth = 0.76.}
    \label{VSIhist}
\end{figure}

\begin{figure}
\centering
\includegraphics[width=0.7\linewidth]{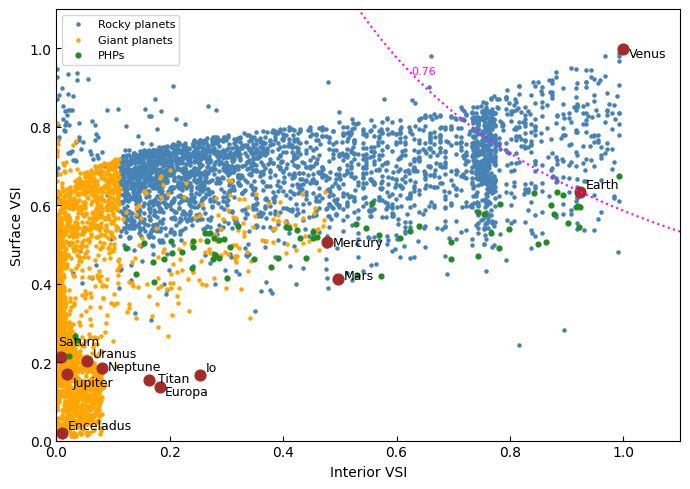}
\caption{Plot of Interior VSI vs Surface VSI. Yellow dots represent \emph{giant planets}, blue dots represent \emph{rocky planets}, brown dots represent \emph{Solar System objects} and green dots represent \emph{Potentially Habitable Planets} (PHPs). The dotted line represents VSI value of Earth (0.76). We have found 341 planets above this threshold.}
\label{VSIscatter}
\end{figure}

The histogram of the global VSI in Fig.~\ref{VSIhist} exhibits a bimodal Gaussian behavior at higher VSI values, in contrast to Fig.~\ref{ESIhist}, where bimodality is observed in the lower-value regime. This difference arises because the majority of exoplanets in current planetary catalogs are characterized by relatively high temperatures. Fig.~\ref{VSIscatter} presents a scatter plot of interior versus surface VSI for the same dataset, with the dashed curve indicating the Earth reference value (VSI = 0.76). Based on this criterion, we identify 341 exoplanets that can be classified as Venus-like, out of which one is a PHP -- Trappist-1e, the fourth planet of the TRAPPIST-1 planetary system. A list of the top 20 closest Venus-like exoplanets with VSI values $>$ 0.76 is presented in Appendix C.

\section{Ancient VSI and Future ESI}\label{sec4.1}

Climate and dynamical modeling indicate that early Venus and Earth may have been remarkably similar in terms of planetary habitability. Both planets could plausibly have supported liquid water oceans, stable land–ocean interfaces, and geochemical environments conducive to prebiotic chemistry \citep{Way2016WasSystem, Way2020VenusianExoplanets}. This parallel early evolution suggests that Venus, like Earth, may once have maintained conditions suitable for biological activity over extended timescales. General circulation models further indicate that Venus could have sustained a temperate surface climate for a period of up to $\sim$2~Gyr prior to the onset of a runaway greenhouse state \citep{Way2020VenusianExoplanets}.

In contrast, present-day Venus represents an extreme and inhospitable environment. Nevertheless, a narrow region within its middle atmosphere, near an altitude of $\sim$55~km, exhibits temperature and pressure conditions comparable to those of Earth’s lower atmosphere \citep{Fegley1997VenusGrinspoon, Patzold2007TheIonosphere, Limaye2018VenusClouds}. The divergence between the ancient and modern states of Venus underscores how planetary habitability can undergo profound transitions in response to changes in key physical and atmospheric parameters. If life did indeed exist on early Venus, some of its representatives could have been transferred to the cloud layer and then adapted to the gradually changing conditions there \citep{Kotsyurbenko2024DifferentVenus}. This points to a connection between ancient and modern Venus and its characterization as a potentially habitable celestial body.

\begin{figure}
\centering
\includegraphics[width=0.7\linewidth]{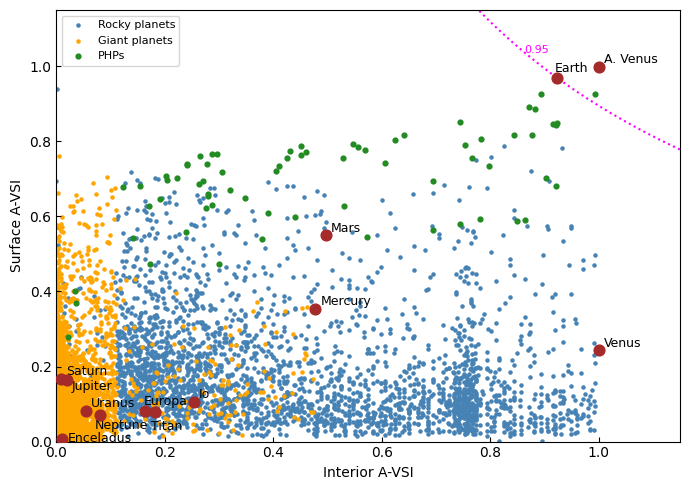}
\caption{Plot of Interior AVSI vs Surface AVSI. Yellow dots represent \emph{giant planets}, blue dots represent \emph{rocky planets}, brown dots represent \emph{Solar System objects} and green dots represent \emph{Potentially Habitable Planets} (PHPs). The dotted line represents VSI value of Earth (0.95). We have found 1 planet beyond this threshold -- Trappist-1e planet.}
\label{AVSI}
\end{figure}

Within this framework, we introduce the Ancient Venus Similarity Index, which provides a quantitative measure of early Earth-like conditions on Venus. We evaluate both interior and surface properties under ancient boundary constraints, adopting the same mathematical formalism described in Section~\ref{sec3}.
The \emph{interior AVSI} is defined as the geometric mean of the similarity components associated with planetary radius and bulk density:
\begin{equation}
AVSI_{\mathrm{interior}} = \left( AVSI_R \times AVSI_\rho \right)^{1/2}.
\end{equation}
The \emph{surface AVSI} is defined as the geometric mean of the similarity components associated with escape velocity and surface temperature:
\begin{equation}
AVSI_{\mathrm{surface}} = \left( AVSI_{V_e} \times AVSI_{T_s} \right)^{1/2}.
\end{equation}
The global AVSI can then be expressed in terms of these two components as
\begin{equation}
AVSI = \left( AVSI_{\mathrm{interior}} \times AVSI_{\mathrm{surface}} \right)^{1/2}.
\end{equation}
The ranges used for weight exponent calculations are identical to those employed previously for all parameters except surface temperature, for which a range of 256–308~K with a reference value of 288~K is adopted, consistent with early Venus climate analogues \citep{Way2016WasSystem}. The AVSI framework enables a comparative assessment of planets—including Venus, Earth, and exoplanets—within a habitability space representative of conditions prior to major climatic divergence. Fig.~\ref{AVSI} presents a scatter plot of interior versus surface AVSI, the dashed curve indicating the reference value of Earth (AVSI = 0.95), implying a 95\% similarity between Venus and Earth under ancient parametric conditions.

\begin{figure}
\centering
\includegraphics[width=0.7\linewidth]{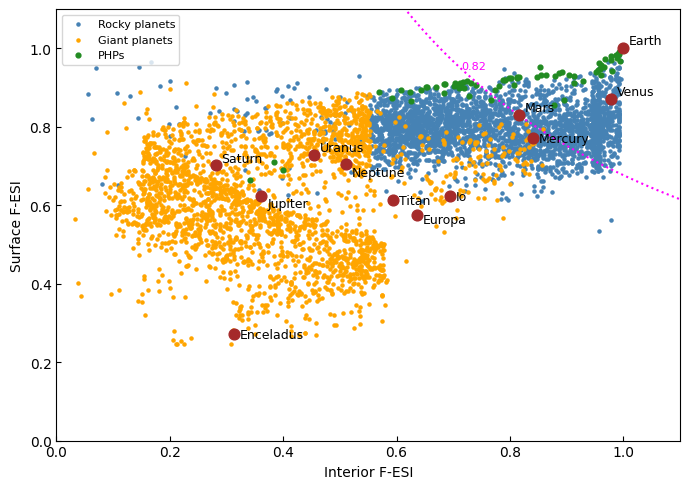}
\caption{Plot of Interior FESI vs Surface FESI. Yellow dots represent \emph{giant planets}, blue dots represent \emph{rocky planets}, brown dots represent \emph{Solar System objects} and green dots represent \emph{Potentially Habitable Planets} (PHPs). The dotted line represents ESI value of Mars (0.82). We have found 1015 planets beyond this threshold.}
\label{FESI}
\end{figure}

Now, let us discuss the habitability transition from ancient Venus to future Earth. The long-term fate of Earth remains a central problem in planetary science and stellar astrophysics. While near-term threats to Earth’s habitability operate on timescales of millions to hundreds of millions of years, the planet’s ultimate destiny is governed by the evolutionary path of the Sun. In approximately 5~Gyr, the Sun is expected to exhaust its core hydrogen and enter the red-giant phase, undergoing substantial radial expansion. Stellar evolution models predict that this expansion will likely engulf the inner planets Mercury and Venus, while Earth’s fate remains uncertain and depends sensitively on factors such as solar mass-loss rates and the efficiency of tidal orbital decay \citep{Rybicki2001OnSystem}. Current models allow outcomes ranging from complete engulfment to survival in a wider but thermally hostile orbit, emphasizing the need for improved constraints on late-stage stellar evolution \citep{Schroder2008DistantRevisited}. Regardless of the outcome, Earth’s future physical state will differ fundamentally from its present conditions.

In this context, we define the Future Earth Similarity Index (FESI), which quantifies the similarity of exoplanets to the projected future state of Earth under late-stage of solar evolution. Following the same mathematical framework as the ESI, 
the \emph{interior FESI} is defined as the geometric mean of the similarity components associated with planetary radius and bulk density:
\begin{equation}
FESI_{\mathrm{interior}} = \left( FESI_R \times FESI_\rho \right)^{1/2}.
\end{equation}

The \emph{surface FESI} is defined as the geometric mean of the similarity components associated with escape velocity and surface temperature:
\begin{equation}
FESI_{\mathrm{surface}} = \left( FESI_{V_e} \times FESI_{T_s} \right)^{1/2}.
\end{equation}

The global FESI can then be expressed in terms of these two components as
\begin{equation}
FESI = \left( FESI_{\mathrm{interior}} \times FESI_{\mathrm{surface}} \right)^{1/2}.
\end{equation}
Within the FESI framework, we adopt radius, density, and escape velocity ranges of 0.5–1.09~EU, 0.7–1.5~EU, and 0.4–1.4~EU, respectively. 
For the surface temperature, we define the lower and upper limits in Eq.~(2) as 288~K and 1603~K, respectively. The lower limit corresponds to the present-day mean surface temperature of Earth, while the upper limit represents the surface temperature expected during the red-giant phase. Within this temperature range, we adopt 737~K as the reference temperature (\(x_0\)), corresponding to the current mean surface temperature of Venus \citep{Way2016WasSystem}.

The resulting weight exponent values are listed in Table~\ref{tab:esivsi}. Fig.~\ref{FESI} shows a scatter plot of interior versus surface FESI, with the dashed curve representing the Mars reference value (FESI$_{\rm Mars}=0.82$). We have found 1015 planets beyond this threshold, of which there are 973 rocky planets, 41 PHPs, and 1 gas giant, identified as TOI-1266 b, a `super Earth' exoplanet.

Among these, 973 are classified as rocky planets, 41 as potentially habitable planets (PHPs), and one, TOI-1266 b, as a gas-rich planet. TOI-1266 b is a low-density sub-Neptune with a hydrogen-rich envelope, making its FESI classification particularly noteworthy. It is important to emphasize, however, that the FESI value represents a snapshot characterization based on the planet's currently observed physical properties and does not account for its temporal evolution. Processes such as atmospheric escape, cooling, contraction, and irradiation-driven mass loss may modify the planet's radius, density, escape velocity, and atmospheric properties over time, potentially altering its FESI. Thus, the present FESI value should be interpreted as a measure of the similarity between the current state of TOI-1266 b and the adopted future-Earth reference state, rather than as an indication that the planet will retain the same FESI throughout its evolution. A quantitative assessment of its future FESI would require coupled planetary-structure and atmospheric-evolution models, which is beyond the scope of the present study.

\begin{table}
\centering
\caption{Future ESI and ancient VSI parametric table }
\label{tab:esivsi}
\small
\begin{tabular}{lcccccc}
\hline
Planetary property & 
\begin{tabular}[c]{@{}c@{}}Ref. value\\for FESI\end{tabular} &
\begin{tabular}[c]{@{}c@{}}Ref. value\\for AVSI\end{tabular} &
\begin{tabular}[c]{@{}c@{}}Weight exp.\\for FESI\end{tabular} &
\begin{tabular}[c]{@{}c@{}}Weight exp.\\for AVSI\end{tabular} \\
\hline
Mean radius & 1 EU & 1 VU  & 0.57 & 2.55 \\
Bulk density & 1 EU & 1 VU  & 1.07 & 3.61 \\
Escape velocity & 1 EU & 1 VU  & 0.70 & 1.71 \\
Surface temperature & 288 K & 288 K & 0.43 & 4.09 \\
\hline
\end{tabular}
\end{table}

In summary, the updated ESI and MSI metrics identify 103 Earth-like and 11 Mars-like exoplanets, while the newly proposed VSI, AVSI, and FESI frameworks classify 341 Venus-like, 1 ancient Venus-like, and 1015 future Earth-like exoplanets, respectively. These results significantly broaden the scope of comparative planetology by demonstrating that potentially habitable or evolutionarily analogous worlds extend far beyond strict Earth analogs, highlighting Venus-like and future Earth-like pathways as dominant evolutionary outcomes and offering a more comprehensive framework for prioritizing targets in atmospheric characterization and habitability studies.

\section{Discussion and Conclusions}\label{sec7}

The concept of planetary habitability, although traditionally associated with the circumstellar habitable zone (HZ), extends well beyond orbital distance from a host star. The HZ delineates the region where liquid water could, in principle, exist on a planetary surface; however, true habitability is additionally governed by factors such as planetary age, atmospheric composition, geochemical cycling, and stellar activity \citep{Kopparapu2014HabitableMass, Safonova2016AgeHabitability, Izenberg2026}. Consequently, a planet may reside within the HZ yet remain inhospitable to life. Venus provides a compelling example: despite orbiting near the inner edge of the Sun’s HZ, it experienced a runaway greenhouse transition that rendered its surface uninhabitable \citep{Kane2024VenusHabitability}. In contrast, Mars, located near the outer boundary of the HZ, preserves geomorphological and geochemical evidence consistent with episodic liquid water and the possibility of ancient microbial habitability \citep{Batalha2014ExploringMission, Cockell2016Habitability:Review}.

From a practical standpoint, the search for habitable worlds has historically focused on identifying Earth-like planets, as Earth remains the only known example of a life-bearing world. Astrobiology, however, seeks to broaden this perspective by considering life in diverse and extreme environments. In this context, attention has increasingly turned toward organisms capable of surviving under conditions far removed from terrestrial norms, including those potentially present on Venus. Venusian science has emphasized the planet’s divergent evolutionary path—from early Earth-like conditions to the extreme surface environment observed today \citep{Way2016WasSystem, Izenberg2021TheEquation, Westall2023TheVenus}. Although the Venusian surface is clearly inhospitable, its cloud layers and middle atmosphere have been proposed as potential niches where extremophile life might persist \citep{Sagan1967LifeVenus, Fegley1997VenusGrinspoon, Limaye2018VenusClouds, Mogul2021VenusClouds}. Dedicated efforts such as the Venus Life Finder (VLF) missions aim to assess the habitability of these atmospheric regions through in situ measurements \citep{Seager2022VenusSummary, Ligterink2022TheMission}. Reanalyses of data from earlier missions, including VENERA, have even identified hypothetical living objects as possible biosignatures \citep{Ksanfomality2019HypotheticalExperiments}, while the reported detection of phosphine in the upper atmosphere has further stimulated debate about extant life on Venus \citep{Greaves2021PhosphineVenus}. These developments underscore the necessity of estimating planetary habitability using observable and measurable properties rather than orbital location alone.

In this work, we have revisited planetary similarity metrics through updated calculations of the Earth Similarity Index and Mars Similarity Index, and by introducing a new Venus Similarity Index. The contrasting evolutionary and habitability results of Earth, Mars and Venus demonstrate that habitability cannot be inferred solely from a planet’s position within the HZ, but must instead be evaluated using multi-parameter frameworks such as the ESI, MSI, and VSI. Our results reveal distinct statistical behaviors for each index, thus providing complementary perspectives on planetary environments. The recalculated ESI with current data exhibits an approximately bimodal distribution skewed toward lower similarity values, indicating that truly Earth-like planets remain statistically rare even within nominal HZs \citep{Safonova2016AgeHabitability}. 
The MSI, by contrast, follows an approximately Gaussian distribution centered near its mean, with most planets occupying intermediate similarity values. This regime corresponds to a domain of “marginal habitability”, where transient surface or subsurface life may be possible, as was hypothesized for early Mars \citep{Cockell2016Habitability:Review}. The VSI introduced here extends this comparative framework toward the hotter, inner regions of the HZ, capturing planets with physical and thermodynamic properties analogous to Venus. The observed bimodality in the VSI distribution suggests the existence of two Venus-like populations: planets undergoing active greenhouse transitions and those residing in partially moderated states. This behavior reinforces the importance of including Venus as a reference case in comparative habitability studies, consistent with its role as an evolutionary model for terrestrial planets \citep{Kane2024VenusHabitability}.

The implications of these findings extend beyond exoplanet classification. The updated ESI and MSI reaffirm established concepts associated with Earth-like and Mars-like environments \citep{Cockell2016Habitability:Review, Schulze-Makuch2011AExoplanets}, while the VSI highlights the distinctiveness and prevalence of Venus analogs.
Following the formal definition of the 'Venus Zone' \citep{Kane2014OnData}, our VSI metric can be used to further classify exoplanets within this region.
 Understanding the relative distributions of Earth-, Mars-, and Venus-like exoplanets informs not only the search for life elsewhere but also the broader narrative of planetary evolution, including the possible futures of terrestrial worlds. As noted by \citet{Safonova2016AgeHabitability}, the study of habitable exoplanets serves two purposes: identifying life beyond Earth and constraining the evolutionary limits of Earth-like planets.

The combined application of the ESI, MSI, and VSI thus provides a unified framework for classifying planetary environments and refining habitability criteria beyond the classical HZ paradigm. By integrating multiple physical parameters, these indices offer a more comprehensive and sustainable approach to planetary characterization. Future studies should extend the VSI to larger and more diverse exoplanet catalogs, particularly those expected from upcoming missions such as PLATO and the Roman Space Telescope, to further evaluate the predictive role of Venus analogs in defining planetary habitability \citep{Turbet2017ClimatePlanets}. 

The AVSI results indicate that very few Venus-like evolutionary states occupy regions close to moderate habitability conditions (say, near the Earth in Fig.~8), suggesting that early Venus may have maintained surface environments capable of supporting liquid water before the onset of runaway greenhouse evolution. The FESI analysis further demonstrates that increasing stellar flux and long-term atmospheric evolution progressively shift terrestrial planets away from the present-day Earth habitability regime, highlighting the temporal instability of habitable conditions. Together, the AVSI and FESI trends shown in the parameter-space distributions support the idea that planetary habitability is not static, but evolves continuously through transitional climatic phases governed by stellar evolution, atmospheric loss, and surface temperature changes. 

Future studies will extend the VSI framework to larger and more diverse exoplanet catalogs from upcoming missions such as PLATO and the Roman Space Telescope, enabling improved statistical constraints on Venus analog populations. Incorporating atmospheric composition, stellar activity, and temporal evolution into the similarity indices will further refine their predictive power for assessing planetary habitability beyond the classical habitable zone.

\section*{Appendix A: ESI, MSI, and VSI parametric table}

\begin{table}[h!] 
\label{tab:esimsivsi}
\small
\begin{tabular}{lcccccc}
\hline
Planetary property & 
\begin{tabular}[c]{@{}c@{}}Ref. value\\for ESI\end{tabular} &
\begin{tabular}[c]{@{}c@{}}Ref. value\\for MSI\end{tabular} &
\begin{tabular}[c]{@{}c@{}}Ref. value\\for VSI\end{tabular} &
\begin{tabular}[c]{@{}c@{}}Weight exp.\\for ESI\end{tabular} &
\begin{tabular}[c]{@{}c@{}}Weight exp.\\for MSI\end{tabular} &
\begin{tabular}[c]{@{}c@{}}Weight exp.\\for VSI\end{tabular} \\
\hline
Mean radius & 1 EU  & 1 MU  & 1 VU  & 0.57 & 0.86 & 2.55 \\
Bulk density & 1 EU  & 1 MU  & 1 VU  & 1.07 & 2.10 & 3.61 \\
Escape velocity & 1 EU  & 1 MU  & 1 VU  & 0.70 & 1.09 & 1.71 \\
Surface temperature & 288 K & 240 K & 737 K & 5.58 & 3.23 & 1.47 \\
\hline
\end{tabular}
\vspace{0.1cm}

\raggedright
\footnotesize{EU = Earth Units, with Earth radius 6371 km, density 5.51 g/cm$^3$, escape velocity 11.19 km/s.\\
MU = Mars Units, with Mars radius 3390 km, density 3.93 g/cm$^3$, escape velocity 5.03 km/s.\\
VU = Venus Units, with Venus radius 6052 km, density 5.24 g/cm$^3$, escape velocity 10.36 km/s.}
\end{table}

\newpage
\section*{Appendix B: Present-day physical properties of Venus}

\begin{table}[h!]
\begin{tabular}{p{4.2cm} p{4.5cm} p{5.2cm}}
\hline
\textbf{Parameter} & \textbf{Value} & \textbf{Reference} \\
\hline
Surface pressure & $\sim$92 bar (9.2 MPa) & \citet{Seiff1985ModelsAltitude}, \citet{Taylor2018Venus:Planet} \\
Global surface temperature & $\sim$735--740 K & \citet{Way2016WasSystem}, \citet{Taylor2018Venus:Planet} \\
Mean solar irradiance & $\sim$2613 W m$^{-2}$ ($\sim$1.9 Earth's) & \citet{Taylor2018Venus:Planet} \\
Solar UV flux & High; strong absorption by SO$_2$ & \citet{Esposito1984SulfurVolcanism}, \citet{Limaye2018VenusClouds} \\
Water vapor (bulk) & $\sim$30 ppm (variable) & \citet{Fegley1997VenusGrinspoon}, \citet{Ignatiev1999WaterSpectra} \\
Gravitational acceleration & 8.87 m s$^{-2}$ (0.9 Earth's) & \citet{Taylor2018Venus:Planet} \\
Escape velocity & 10.36 km s$^{-1}$ & \citet{Taylor2018Venus:Planet} \\
Day length (sidereal) & 243 Earth days (retrograde) & \citet{Taylor2018Venus:Planet} \\
Year length & 224.7 Earth days & \citet{Taylor2018Venus:Planet} \\
Equatorial diameter & 12,104 km (0.95 Earth's) & \citet{Taylor2018Venus:Planet} \\
Mass & $4.87 \times 10^{24}$ kg (0.82 Earth's) & \citet{Taylor2018Venus:Planet} \\
Surface water & None & \citet{Way2016WasSystem} \\ \hline
Atmospheric Composition \\ \hline 
\quad CO$_2$ & 96.5\% & \citet{Fegley1997VenusGrinspoon} \\
\quad N$_2$ & 3.5\% & \citet{Fegley1997VenusGrinspoon} \\
\quad SO$_2$ & $\sim$150 ppm (variable) & \citet{Esposito1984SulfurVolcanism} \\
\quad H$_2$O & $\sim$30 ppm & \citet{Fegley1997VenusGrinspoon} \\
\quad CO & $\sim$17 ppm & \citet{Taylor2018Venus:Planet} \\
\quad Ar & $\sim$70 ppm & \citet{Taylor2018Venus:Planet} \\
Cloud composition & H$_2$SO$_4$ aerosols & \citet{Limaye2018VenusClouds} \\
Lithosphere composition & Basaltic (Fe-rich) & \citet{Fegley1997VenusGrinspoon}, \citet{Taylor2018Venus:Planet}  \\
\hline
\end{tabular}
\end{table}

\newpage
\section*{Appendix C: Top 20 closest exoplanets with VSI$>$0.76}
\begin{table}[h!]
\centering
\begin{tabular}{lccc}
\hline
\textbf{Planet Name} & \textbf{VSI} & \textbf{Distance from Earth (pc)} \\
\hline
 tau Cet g & 0.812 & 3.603 \\
 GJ 1061 b & 0.775 & 3.672 \\
 YZ Cet b & 0.882 & 3.712 \\
 YZ Cet c & 0.827 & 3.712 \\
 YZ Cet d & 0.803 & 3.712 \\
 GJ 273 c & 0.836 & 3.786 \\
 HD 20794 c & 0.768 & 6.002 \\
 GJ 581 e & 0.797 & 6.298 \\
 LTT 1445 A c & 0.813 & 6.869 \\
 GJ 393 b & 0.770 & 7.031 \\
 GJ 357 b & 0.771 & 9.441 \\
 AU Mic d & 0.884 & 9.722 \\
 HD 26065 b & 0.779 & 10.005 \\
 TRAPPIST-1 c & 0.768 & 12.429 \\
 TRAPPIST-1 b & 0.791 & 12.429 \\
 LHS 475 b & 0.939 & 12.481 \\
 GJ 1132 b & 0.779 & 12.613 \\
 TOI-540 b & 0.923 & 14.002 \\
 LHS 3844 b & 0.784 & 14.884 \\
 G 264-012 b & 0.768 & 16.009 \\
\hline
\end{tabular}
\end{table}

\section*{Funding}
This research did not receive any specific grant from funding agencies in the public, commercial, or not-for-profit sectors.

\section*{Conflict of Interest disclosure} 

The authors declare that there are no conflicts of interest for this manuscript.

\section*{Acknowledgment}

The authors thank the PHL team for providing data for this work.

\section*{Data Availability}

The dataset has been archived in a publicly accessible repository: Raviprakash, Swathi; Jagadeesh, Madhu Kashyap; Safonova, Margarita (2026), “Dataset of Habitability Study of Terrestrial Planets: Application to Venus-like Worlds”, Mendeley Data, V1. \\
https://doi.org/10.17632/c28hgfn6cr.1

\end{document}